\documentclass[twocolumn,prd,floatfix,nofootinbib,superscriptaddress,longbibliography]{revtex4-1}

\pdfoutput=1

\usepackage{amsmath, amssymb, amsfonts, amsthm, latexsym, epsfig, mathrsfs, xcolor, bbm, slashed}

\usepackage[inline]{enumitem}

\usepackage{setspace}
\usepackage[marginal, multiple]{footmisc}

\usepackage[T1]{fontenc}
\usepackage[utf8]{inputenc}
\usepackage{lmodern}

\usepackage[unicode, colorlinks, allcolors=blue!70!black, linktocpage, pdfusetitle]{hyperref}
\usepackage[all]{hypcap}

\usepackage{tikz}
\usetikzlibrary{arrows.meta}

\numberwithin{equation}{section}

\usepackage{cleveref}

\usepackage{cancel}

\usepackage{microtype}

\setlist[enumerate]{noitemsep, label=(\arabic*), ref=(\arabic*)}

\newlist{condlist}{enumerate}{2}
\setlist[condlist,1]{noitemsep, label=(\arabic*), ref=(\arabic*)}
\setlist[condlist,2]{noitemsep, label=(\alph*), ref=(\arabic{condlisti}.\alph*)}
\crefname{condlisti}{condition}{conditions}
\crefname{condlistii}{condition}{conditions}

\renewcommand\thesection{\arabic{section}}
\renewcommand\thesubsection{\arabic{subsection}}

\makeatletter
\def\p@subsection{\thesection.}
\def\p@subsubsection{\thesection.\thesubsection.}
\makeatother

\theoremstyle{plain}
\newtheorem{thm}{Theorem}

\theoremstyle{definition}

\theoremstyle{remark}

\crefname{equation}{Eq.}{Eqs.}
\creflabelformat{equation}{#2#1#3}

\crefname{section}{\S}{\S}
\crefname{appendix}{Appendix}{Appendices}
\crefname{figure}{Fig.}{Figs.}

\crefname{definition}{Def.}{Defs.}
\crefname{prop}{Prop.}{Props.}
\crefname{lemma}{Lemma}{Lemmas}
\crefname{corollary}{Cor.}{Cors.}
\crefname{thm}{Theorem}{Theorems}
\crefname{remark}{Remark}{Remarks}

\crefname{ass}{Assumptions}{Assumptions}
\crefname{property}{Properties}{Properties}

\newcommand{\be}{\begin{equation}}
\newcommand{\ee}{\end{equation}}
\newcommand{\nn}{\nonumber}

\newcommand{\lb}{\left}
\newcommand{\rb}{\right}

\newcommand{\mc}{\mathcal}

\newcommand{\ms}{\mathscr}

\newcommand{\bb}{\mathbb}

\newcommand{\eqsp}{\, ,\quad}

\newcommand{\hr}{\begin{center}* * *\end{center}}

\newcommand{\union}{\cup}
\newcommand{\inter}{\cap}

\newcommand{\Lie}{\pounds}
\newcommand{\hatLie}{\Lie\kern-0.25em\hat{\vphantom{\Lie{}}}\kern0.25em}
\newcommand{\defn}{\mathrel{\mathop:}=}

\newcommand{\pd}{\partial}
\newcommand{\cd}{\nabla}

\newcommand{\df}[1]{\boldsymbol{#1}}

\newcommand{\af}[1]{\mathring{#1}}

\begin{document}

\title{Black hole scalar charge from a topological horizon integral \\
    in Einstein-dilaton-Gauss-Bonnet gravity}

\author{Kartik Prabhu}\email{kartikprabhu@cornell.edu}
\affiliation{%
  Cornell Laboratory for Accelerator-based Sciences and Education (CLASSE),\\
  Cornell University, Ithaca, NY 14853, USA}

\author{Leo C.~Stein}
\email{leostein@tapir.caltech.edu}

\affiliation{%
  TAPIR, Walter Burke Institute for Theoretical Physics,
  California Institute of Technology, Pasadena, California 91125, USA}

\hypersetup{pdfauthor={Prabhu and Stein}}

\begin{abstract}
  In theories of gravity that include a scalar field, a compact object's scalar charge is a crucial quantity since it controls dipole
  radiation, which can be strongly constrained by pulsar timing and
  gravitational wave observations.
  However in most such theories, computing the scalar
  charge requires simultaneously solving the coupled, nonlinear metric and
  scalar field equations of motion.
  In this article we prove that in linearly-coupled
  Einstein-dilaton-Gauss-Bonnet gravity, a black hole's scalar charge
  is completely determined by the horizon surface gravity times the
  Euler characteristic of the bifurcation surface, without solving any
  equations of motion.
  Within this theory, black holes announce their horizon topology and
  surface gravity to the rest of the universe through the dilaton
  field.
  In our proof, a 4-dimensional topological density descends to a
  2-dimensional topological density on the bifurcation surface of
  a Killing horizon.
  We also comment on how our proof can be generalized to other
  topological densities on general \(G\)-bundles, and to theories
  where the dilaton is non-linearly coupled to the Euler density.
\end{abstract}

\maketitle

\section{Introduction}\label{sec:intro}

Despite the theoretical beauty and continued consistency with
observations~\cite{Will:2014kxa} of general relativity (GR), there are
strong motivations for studying theories of gravity beyond GR.  These
range from attempts at quantum theories of
gravity~\cite{Rovelli:1997qj, Barrau:2017tcd}, to trying to explain
some phenomenon or fix some problem (inflation~\cite{Cheung:2007st,
  Weinberg:2008hq}, dark matter~\cite{Famaey:2011kh, Clifton:2011jh},
dark energy~\cite{Clifton:2011jh, Joyce:2016vqv}, black hole
information~\cite{Unruh:2017uaw, Marolf:2017jkr}) by changing the
gravity theory, or exploring the theory space to better understand
gravity theories.

Almost all beyond-GR theories include additional degrees of freedom
and a large number of proposed beyond-GR theories include one or more
massless or very light scalar fields~\cite{Sotiriou:2015lxa}.  With a long-ranged
scalar field in the theory, compact objects (e.g.~black holes or
neutron stars) may acquire a \emph{scalar charge}: the
spherically-symmetric, $1/r$ component of the scalar field sourced by
a body.  Scalar charges are crucial in the dynamics of a
compact-object binary system, as they control the presence or absence
of scalar dipole radiation.  The presence of dipole radiation would
dominate over the otherwise-leading quadrupolar emission of
gravitational waves, a so-called ``pre-Newtonian'' correction.  Such
an effect can be strongly constrained by both pulsar timing and the
direct detection of gravitational waves~\cite{Berti:2015itd}.

In most theories with scalar fields, computing a black hole's scalar
charge requires solving the coupled set of metric and scalar field
equations in the nonlinear gravity regime.  On occasion, this can be
accomplished analytically with symmetry reduction
\cite{Garfinkle:1990qj, Kanti:1995vq}, and/or simplifying assumptions such as
a perturbative treatment away from GR \cite{Mignemi:1992nt}.
However, the general case requires numerics for fully nonlinear
partial differential equations \cite{Kleihaus:2015aje}.

A dramatic simplification occurs in linearly-coupled
Einstein-dilaton-Gauss-Bonnet gravity (EdGB). 
Already in~\cite{Yunes:2016jcc,Berti:2018cxi}, the authors had found
through explicit calculations, perturbative in
$\alpha$, a coupling parameter, that the dilaton charge $q$ on the Kerr background is
(in our conventions)
\begin{align}
  \label{eq:charge-Kerr-1}
    q = \alpha\frac{a^{2}-M^{2}+M\sqrt{M^{2}-a^{2}}}{2Ma^{2}}
  + \mathcal{O}(\alpha^{2})
  \,.
\end{align}
However it was not noticed that this particular
combination of mass and spin is equal to the Kerr surface gravity
(Eq.~12.5.4 of \cite{Wald-book}), so that in fact
$q = \alpha\kappa_{\text{Kerr}} + \mathcal{O}(\alpha^{2})$.

 We prove (\cref{thm:main-thm}) that in this
theory the dilaton scalar charge is given by the rather simple expression 
\begin{align}
  \label{eq:intro-result}
  q = \tfrac{1}{2} \alpha \kappa \text{Euler}(B) \,,
\end{align}
where $\kappa$ is the surface
gravity of a Killing horizon, and Euler$(B)$ is the 
Euler characteristic of the bifurcation surface.
Our proof is valid to all orders in $\alpha$,
and moreover does not require explicitly solving any field
equations.
In fact the metric does not need to satisfy any equations
of motion; the proof is valid on any asymptotically flat,
stationary-axisymmetric black hole spacetime.
Thus in linearly-coupled EdGB, the horizon topology and
surface gravity of black holes are known at spatial infinity by
looking at the asymptotic falloff of the dilaton.

We present the proof of our main result in \cref{sec:proof}, which relies on a certain divergence identity (\cref{eq:J-Q-defn}) for a massless scalar field linearly coupled to a topological
form and the Killing symmetry of the black hole spacetime.  Combining these ingredients allows the 4-dimensional Euler (or Gauss-Bonnet)
topological density to descend to the 2-dimensional Euler
characteristic on the bifurcation surface.
In \cref{sec:disc}, we discuss the implications of this
proof, and how it can be generalised to theories where the dilaton is nonlinearly-coupled to the Euler density. We also show how to extend our proof to higher dimensions and other
invariants, such as Chern characters of complex $G$-bundles.  We demonstrate this with an example of an axion coupled to
electromagnetism, wherein the axion charge measures the
electromagnetic potential of the horizon times the magnetic
monopole charge of the black hole.

\section*{Notation and conventions}
We follow the factor and sign conventions in \cite{Wald-book}. Tensor fields on spacetime will be denoted by abstract indices
\(\mu,\nu,\lambda,\ldots\) from the lowercase Greek
alphabet, and we use lowercase letters from the beginning of the Latin
alphabet $a,b,c,\ldots$ to denote tensors on the orthonormal frame
bundle, summarized in \cref{sec:tetrads}.
When using an index-free notation
for differential forms and vector fields we denote differential forms
by a bold-face symbol. When translating differential forms to and from an
index notation, we use the symbol \(\equiv\) to denote such a translation, for example, for a \(p\)-form \(\df A\) we have \((\df A)_{\mu_1 \ldots \mu_p} \equiv A_{\mu_1 \ldots \mu_p}\). Our conventions for the \emph{volume form}
\(\df\varepsilon_4\),
the \emph{Hodge dual} \(*\) and the \emph{interior product} are as follows
\begin{subequations}\begin{align}
     \df\varepsilon_4 & \equiv \varepsilon_{\mu_1 \ldots \mu_4} \\
    \varepsilon^{\mu_1 \ldots \mu_k \mu_{k+1} \ldots \mu_4} \varepsilon_{\mu_1 \ldots \mu_k \nu_{k+1} \ldots \nu_4} & = - (4-k)! k! \delta^{[\mu_{k+1}}_{\nu_{k+1}} \cdots \delta^{\mu_4]}_{\nu_4} \\
    (* \df A)_{\mu_1 \ldots \mu_{4-p}} & \equiv \tfrac{1}{p!} \varepsilon^{\nu_1 \ldots \nu_p}{}_{\mu_1 \ldots \mu_{4-p}} A_{\nu_1 \ldots \nu_p} \\
    (X \cdot \df A)_{\mu_1 \ldots \mu_{p-1}} & \equiv X^\nu A_{\nu \mu_1 \ldots \mu_{p-1}}
\end{align}\end{subequations}
where \(\df A\) is a \(p\)-form and \(X^\mu\) is some vector field.

\section{Lagrangian and the dilaton charge}
\label{sec:proof}

We consider a theory with gravity on a 4-dimensional spacetime \(M\) with a Lorentzian metric \(g_{\mu\nu}\) and a scalar dilaton field \(\vartheta\). The dynamics of the theory is given by the Lagrangian \(4\)-form
\be\begin{split}
    \df L & = \df L_{gravity} + \df L_{\vartheta} \\
\end{split}\ee
where \(\df L_{gravity}\) is some gravitational Lagrangian which is independent of the dilaton \(\vartheta\).

There are several theories in the literature which are referred to as
\emph{Einstein-dilaton-Gauss-Bonnet}, commonly with an exponential
coupling~\cite{Metsaev:1987zx,Maeda:2009uy} between a dilaton and the
Euler density (defined below). For our main result we will consider a
linear coupling (which admits a shift symmetry $\vartheta \to
\vartheta + \mathrm{const.}$),
and comment on the extension to more general couplings in \cref{sec:disc}.
We take the dilaton-Gauss-Bonnet Lagrangian \(\df L_\vartheta\) to be
\be\label{eq:L-dilaton}\begin{split}
    \df L_\vartheta & = \tfrac{1}{2} (* d\vartheta) \wedge d\vartheta + \tfrac{\alpha}{8} \vartheta \df{\mc E} \\
    & = \df\varepsilon_4 \lb( -\tfrac{1}{2} \nabla_\mu\vartheta \nabla^\mu \vartheta + \tfrac{\alpha}{8} \vartheta \mc E \rb)
\end{split}\ee
where the \(4\)-form \(\df{\mc E}\) corresponds to the 4-dimensional
\emph{Euler density} as
\be\label{eq:Euler-density}\begin{split}
    \df{\mc E} & = \df\varepsilon_4~ \mc E \\
    \text{with}\quad \mc E & = - ({}^*R^*)^{\mu\nu\lambda\rho} R_{\mu\nu\lambda\rho} \\
    & = R_{\mu\nu\lambda\rho} R^{\mu\nu\lambda\rho} - 4 R_{\mu\nu} R^{\mu\nu} + R^2 
    \,,
\end{split}\ee 
and the double dual of the Riemann tensor is defined as
\be\label{eq:dd-Riem}
    ({}^*R^*)^{\mu\nu\lambda\rho} \defn \tfrac{1}{4} \varepsilon^{\mu\nu\sigma\tau} R_{\sigma\tau\gamma\delta} \varepsilon^{\gamma\delta \lambda\rho}
\,.
\ee 

While our final result can be presented in terms of tensor fields on spacetime, it will be, instead, convenient to use orthonormal tetrads \(\df e^a\) and a connection \(\df\omega^a{}_b\). It is more natural in the following analysis to treat \(\df e^a\) and \(\df\omega^a{}_b\) as \emph{globally} well-defined fields on a \emph{principal bundle}, following the treatment in \cite{KP-first-law} (see also \cite{JM})---we summarise the essential points in \cref{sec:tetrads}. Readers unfamiliar with the bundle formalism can skip to \cref{eq:Q-tensor} where we present the tensorial form of the charge used in our main result \cref{thm:main-thm}.

In terms of the connection \(\df\omega^a{}_b\), we can write \(\df{\mc E}\) as an exact form \cite{Nieh-GB,CRGV,KP-first-law}
\be\label{eq:Euler}\begin{split}
    \df{\mc E} & = \epsilon_{abcd} \df R^{ab} \wedge \df R^{cd} = d \df\Upsilon \\
    \text{ with }
    \df\Upsilon & = \epsilon_{abcd} \df \omega^{ab} \wedge \lb( \df R^{cd} - \tfrac{1}{3} {\df \omega^c}_e \wedge \df \omega^{ed} \rb) 
    \,,
\end{split}\ee
where \(\df R^a{}_b\) is the curvature \(2\)-form of the connection (see \cref{eq:R-defn}). We emphasise that the \(3\)-form \(\df\Upsilon\) \emph{cannot} be represented as a covariant tensor on spacetime---even if one writes \(\df\Upsilon\) in some local coordinate system it necessarily involves undifferentiated Christoffel symbols which are not covariant tensors. In the language of principal bundles, while \(\df\Upsilon\) is globally well-defined, it is not a horizontal form on the bundle.

Varying the Lagrangian with the dilaton \(\vartheta\) gives the dilaton equation of motion
\be\label{eq:dilaton-eom}\begin{split}
    0 = \df E_\vartheta = d * d\vartheta + \tfrac{\alpha}{8} \df{\mc E} = \df\varepsilon_4 \lb( \square \vartheta + \tfrac{\alpha}{8} \mc E \rb)
\end{split}\ee
where \(\square \defn \nabla_\mu \nabla^\mu\) is the wave operator. Note that we \emph{do not} impose the gravitational
equations of motion obtained by varying the metric \(g_{\mu\nu}\), that is the metric can be considered as a ``background field''.

Using \cref{eq:Euler} the dilaton equation of motion can be written as an exact form, and one could attempt to integrate this over some region of spacetime bounded by two Cauchy surfaces to get the scalar charge. However, since \(\df\Upsilon\) is not covariant (as explained above), the result would depend on the choice of coordinate system, or equivalently on the choice of orthonormal tetrads.

To avoid using non-covariant quantities, we proceed instead as follows. Using \cref{eq:Lie-conn} we have, for any vector field \(X^\mu\)
\be\begin{split}
    \epsilon_{abcd} \df R^{ab} \wedge \Lie_X \df\omega^{cd}
    & = \epsilon_{abcd} \df R^{ab} \wedge D(X \cdot \df\omega^{cd}) + \tfrac{1}{2} X \cdot \df{\mc E} \\
    & = d \lb[ \epsilon_{abcd} \df R^{ab} (X \cdot \df\omega^{cd}) \rb] + \tfrac{1}{2} X \cdot \df{\mc E} 
\end{split}\ee
Using this in \cref{eq:dilaton-eom}, we get the divergence identity
\begin{subequations}\label{eq:J-Q-defn}\begin{align}
    \begin{split}
    \df{\mc J}_X & = d \df{\mc Q}_X + X \cdot \df E_\vartheta \\
    & = \Lie_X (*d\vartheta) + \tfrac{\alpha}{4} \epsilon_{abcd} \df R^{ab} \wedge \Lie_X \df\omega^{cd} 
    \end{split} \label{eq:J-dQ} \\
    \text{ where }
    \df{\mc Q}_X & = X \cdot (*d\vartheta) + \tfrac{\alpha}{4} \epsilon_{abcd} \df R^{ab} (X \cdot \df\omega^{cd}) \label{eq:Q-defn}
\end{align}\end{subequations} 
The current \(\df{\mc J}_X\) is conserved for any vector field
\(X^\mu\), i.e., \(d\df{\mc J}_X = 0\), on solutions to the dilaton
equation of motion. Further, \(\df{\mc J}_X = 0\) whenever \(X^\mu\)
is a symmetry i.e.~\(\Lie_X g_{\mu\nu} = \Lie_X \vartheta = 0\).
\(\df{\mc J}_X\) vanishes on symmetries because it is related to
the Lie derivative of the Noether current for the shift symmetry
$\vartheta \to \vartheta + \mathrm{const}$.
In this case,
the expression for the charge can be simplified using \cref{eq:symm-cond} to get
\be\label{eq:charge-symm}
    \df{\mc Q}_X = X \cdot (*d\vartheta) - \tfrac{\alpha}{4} \epsilon_{abcd} \df R^{ab} ( e^c_\mu e^d_\nu \cd^\mu X^\nu)
\,.
\ee
We also provide a tensorial expression as follows. Define \(\mc Q^{\mu\nu}_X\) from \(\df{\mc Q}_X\) by
\be
    (\df{\mc Q}_X)_{\mu\nu} \equiv \mc Q^{\lambda\rho}_X \varepsilon_{\lambda\rho\mu\nu}
\ee
so that the current vector \(\mc J^\mu_X\) is given by
\be
\begin{split}
(\df{\mc J}_X)_{\mu\nu\lambda} &\equiv \mc J^\rho_X \varepsilon_{\rho\mu\nu\lambda} \\
\text{with}\quad\qquad \mc J^{\mu}_X &= \nabla_\nu \mc Q^{\mu\nu}_X + X^\mu (\square \vartheta + \tfrac{\alpha}{8} \mc E)
\,.
\end{split}
\ee
For a symmetry \(X^\mu\) we can compute (using \cref{eq:charge-symm})
\be
\label{eq:Q-tensor}
\mc Q^{\mu\nu}_X = - 2X^{[\mu}\nabla^{\nu]} \vartheta + \tfrac{\alpha}{2} ({}^*R^*)^{\mu\nu\lambda\rho} \nabla_\lambda X_\rho
\,.
\ee
It can be checked that 
\(\mc J^{\mu}_X = 0\), using the identities
\(\nabla_{\mu} ({}^*R^*)^{\mu\nu\lambda\rho} = 0\) (which follows from
the Bianchi identity \cref{eq:bianchi}), \(\nabla_\mu \nabla_\nu
X_\lambda = R_{\lambda\nu\mu}{}^\rho X_\rho\) for a Killing field
\(X^\mu\) (see Eq.~C.3.6 \cite{Wald-book}), and \(C^{\mu\lambda\rho\sigma} C_{\nu\lambda\rho\sigma} = \tfrac{1}{4} \delta^\mu_\nu C^{\tau\lambda\rho\sigma} C_{\tau\lambda\rho\sigma}\) for the \emph{Weyl tensor} \(C_{\mu\nu\lambda\rho}\) in \(4\)-dimensions \cite{Edgar:2001vv}. The result of
\cref{thm:main-thm} can be obtained by evaluating the integral \(0 =
\int_\Sigma u_\mu \mc J^\mu_K\) where, \(K^\mu\) is the horizon
Killing field and \(u^\mu\) is the future-pointing unit time-like
normal to a Cauchy surface \(\Sigma\), as described below.

\hr

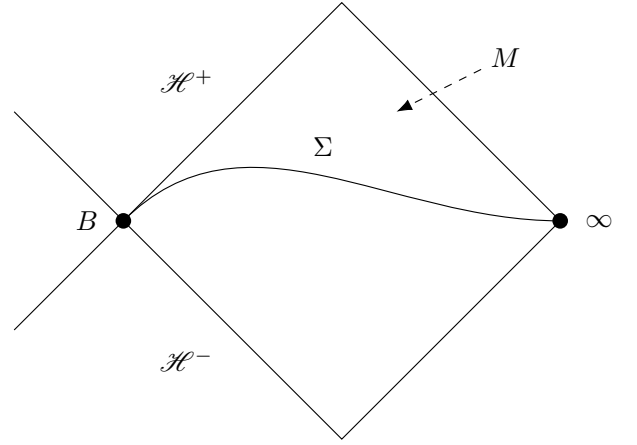
\begin{figure}[tb]
  \begin{center}
    \def\spacetimestandalone{}
    \ifx\spacetimestandalone\undefined
    \begin{tikzpicture}
  [scale=1.4,
  thept/.style={circle,fill=black,inner sep=0pt,outer sep=1mm,minimum size=2mm}]

  \coordinate (i0) at (2,0);
  \coordinate (iplus) at (0,2);
  \coordinate (iminus) at (0,-2);
  \coordinate (B) at (-2,0);
  \coordinate (midHlefttop) at (-3,1);
  \coordinate (midHleftbot) at (-3,-1); \coordinate (Mtarget) at (.5,1);

  \node at (B) [thept,label=left:$B$] {};
  \node at (i0) [thept,label=right:$\infty$] {};

  \draw (midHlefttop) --
  (B) to node[label=below left:$\mathscr{H}^{-}$] {} (iminus) -- (i0) --
  (iplus) to node[label=above left:$\mathscr{H}^{+}$] {} (B) -- (midHleftbot);

  \draw (B) to [out=45,in=180] node[label=above:$\Sigma$] {} (i0);

  \node (M) at (1.5,1.5) {$M$};

  \draw[dashed, arrows={-Latex[length=2mm]}] (M) to (Mtarget);

\end{tikzpicture}
    \else
    \includegraphics[width=8.21cm]{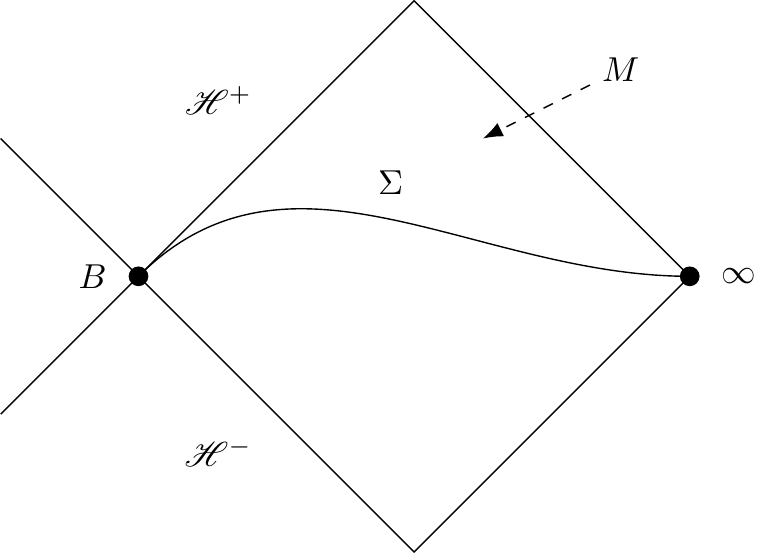}
    \fi
  \end{center}
	\caption{%
    Schematic diagram for black hole spacetime. Note this should not
    be considered as a Carter-Penrose diagram, in particular we make
    no assumptions about the existence of null infinity.
    \label{fig:spacetime}}
\end{figure}

We now consider an asymptotically flat, stationary-axisymmetric black
hole spacetime \((M, g_{\mu\nu})\) shown in \cref{fig:spacetime}, with
a stationary-axisymmetric dilaton field \(\vartheta\) satisfying the
equation of motion \cref{eq:dilaton-eom} (we take all fields to be
smooth ($C^{\infty}$) throughout $M$).  We assume the spacetime has a
bifurcate Killing horizon \(\ms H \defn \ms H^+ \union \ms H^-\), with
a bifurcation surface \(B \defn \ms H^+ \inter \ms H^-\).  We assume
that \(B\) is compact but do not assume other restrictions on its
topology.\footnote{We need not assume that $B$ is connected but, for
  notational convenience, we will assume that this is the case.}  Let
the Killing field generating \(\ms H\) be
\(K^\mu = t^\mu + \Omega_{\ms H} \phi^\mu\) where $t^\mu$ denotes the
time translation Killing field and $\phi^\mu$ denotes the axial
Killing field associated with the horizon rotation parameter
\(\Omega_{\ms H}\).  Let $\Sigma$ denote a Cauchy surface for the
black hole exterior. We assume that $\Sigma$ has one asymptotically
flat end (with asymptotic conditions given by \cref{eq:asymp-falloff}
below), and a boundary at $B$.

The asymptotic flatness conditions on our spacetime are as follows.  There exist asymptotically Minkowskian coordinates \(x^\mu = (t,x,y,z)\) such that
the global Killing fields asymptote to the Minkowski ones at the rates
\be
    t^\mu = (\pd_t)^\mu + \mathcal{O}(1/r) \eqsp \phi^\mu =
    (\pd_\phi)^\mu
    \big( 1+\mathcal{O}(1/r) \big) \,,
\ee
the metric and dilaton asymptote at the rates
\be\label{eq:asymp-falloff}
    g_{\mu\nu} = \af g_{\mu\nu} + \mc O(1/r) \eqsp \vartheta = \vartheta_\infty(\theta) + \frac{q(\theta)}{r} + \mc O(1/r^2)
\ee
where \(\af g_{\mu\nu}dx^{\mu}dx^{\nu} \equiv - dt^2 + dx^2 + dy^2 + dz^2\) is the Minkowskian flat metric in these coordinates, and \((r, \theta, \phi)\) are defined in terms of \((x,y,z)\) in the standard way. In addition, all \(n\)\textsuperscript{th} derivatives of the above quantities (with respect to these coordinates) are required to fall off faster by an additional factor of \(1/r^n\).\footnote{Since we do not impose the gravitational equations of motion we do not need to ensure that such asymptotically flat spacetimes exist as solutions.}

For such spacetimes, we define the \emph{scalar charge} to be the
spherically-symmetric part of the asymptotic $1/r$ falloff (see
also~\cite{Garfinkle:1990qj,YSY})
\be\label{eq:charge-defn}
   q \defn -\frac{1}{4\pi} \int_\infty \df\varepsilon_2 \partial_r \vartheta
   = \frac{1}{4\pi}\int_\infty \df\varepsilon_2 \frac{q(\theta)}{r^2}
   \,.
\ee
Here, and henceforth, \(\int_\infty\) means that the integral is evaluated over an asymptotic \(2\)-sphere \(S_r\) of radius \(r\) and then one takes the limit \(r \to \infty\), and \(\df\varepsilon_2\) is the induced area element  on \(S_r\).\footnote{The induced area element $\df \varepsilon_{2}$ grows as $r^{2}$ and \cref{eq:charge-defn} converges in the limit as $r\to\infty$.}

Now, we use \cref{eq:J-Q-defn} to prove our main result.

\begin{thm}\label{thm:main-thm}
On any asymptotically flat, stationary-axisymmetric black hole spacetime (as defined above) the scalar charge \cref{eq:charge-defn} is given by
\be\label{eq:main}
    q = \tfrac{1}{2} \alpha \kappa {\rm Euler}(B)
    \,,
\ee
where \(\kappa\) is the surface gravity of the black hole and  \({\rm Euler}(B)\) is the \emph{Euler characteristic} of the bifurcation surface \(B\).
\begin{proof}
For \(X^\mu = K^\mu\), the horizon Killing field, integrate \(0 = \df{\mc J}_K = d \df{\mc Q}_K\) on the Cauchy surface \(\Sigma\) to get
\be\label{eq:charge-eqn}
    \int_\infty \df{\mc Q}_K  = \int_B \df{\mc Q}_K
\ee
where we have assumed that the induced orientations in both integrals are outward-pointing i.e.~the space-like normals in \(\Sigma\) point towards \(\infty\). We now evaluate each side of the above expression using \cref{eq:charge-symm}.

To compute the contribution to the charge at infinity, let \(\af{\df e}^a = (dt,dx,dy,dz)\) be an asymptotic tetrad adapted to the asymptotically Minkowskian coordinates. From \cref{eq:asymp-falloff}, we have near infinity
\be\label{eq:tetrad-falloff}
    \df e^a = \af{\df e}^a + \mc O(1/r) \eqsp \df \omega^{ab} = \mc O(1/r^2) \eqsp \df R^{ab} = \mc O(1/r^3)
\,.
\ee
Then, using \cref{eq:charge-defn}, we find
\be\label{eq:charge-inf}
\begin{split}
  \int_\infty \df{\mc Q}_K & = \int_\infty K \cdot (*d\vartheta) = - \int_\infty \df\varepsilon_2 \partial_r \vartheta \\
  &{} = 4\pi q
\,,
\end{split}\ee
where the curvature terms do not contribute due to the falloffs in \cref{eq:tetrad-falloff}, and the contribution from the  \(\phi^\mu\)-part of \(K^\mu\) vanishes since \(\phi^\mu\) is tangent to the spheres at infinity.

On \(B\), we have \(K^\mu\vert_B = 0\) and \(\nabla_\mu K_\nu \vert_B = \kappa \tilde\varepsilon_{\mu \nu} \) where \(\tilde\varepsilon_{\mu\nu}\) is the \emph{binormal} to the bifurcation surface (see \S\ 12.5 \cite{Wald-book}). It can be shown that \(\kappa\) is a constant over \emph{any} bifurcate Killing horizon \(\ms H\) \cite{KayWald}. Using this in \cref{eq:charge-symm}, and noting that \((\df\varepsilon_2)_{\mu\nu} = - \tfrac{1}{2} \varepsilon^{\lambda\rho}{}_{\mu\nu} \tilde\varepsilon_{\lambda\rho}\) is the intrinsic area element to \(B\), we get
\be\label{eq:charge-B}\begin{split}
    \int_B \df{\mc Q}_K & = \tfrac{\alpha}{2} \kappa \int_B \epsilon_{ab} \df R^{ab} = \tfrac{\alpha}{2} \kappa \int_B \df\varepsilon_2 R_2 \\
    & = 2\pi\alpha \kappa {\rm Euler}(B)
\,,
\end{split}\ee
where \(\epsilon_{ab}\) is the tetrad component of
\(\df\varepsilon_2\), \(R_2\) is
the intrinsic Ricci scalar of \(B\), and the last line uses the
2-dimensional Gauss-Bonnet theorem \cite{BT-book}.

Combining \cref{eq:charge-eqn,eq:charge-B,eq:charge-inf}, we have our result \cref{eq:main}.
\end{proof}
\end{thm}

We note here that the contribution to the scalar charge from \(B\) (\cref{eq:charge-B}) can be written as
\be\label{eq:charge-Wald}
    \int_B \df{\mc Q}_K = -\tfrac{\alpha}{8} \kappa \int_B \df\varepsilon_2 \frac{\delta \mc E}{\delta R_{\mu\nu\lambda\rho}} \tilde\varepsilon_{\mu\nu} \tilde\varepsilon_{\lambda\rho}
\ee
in analogy with the \emph{Wald entropy formula} \cite{W-noether-entropy,IW-noether-entropy}. This relation arises due to the second term in \cref{eq:Q-defn}.

\section{Discussion and extension}
\label{sec:disc}

A concise interpretation of this result is that black holes
communicate their horizon topology and surface gravity to spatial
infinity, by encoding this information in the asymptotic falloff of
the dilaton.

When the bifurcation surface \(B\) is a topological \(2\)-sphere, as
is the case for a Kerr spacetime (or any continuous deformation of
Kerr), we have \(q = \alpha \kappa\), consistent with
\cref{eq:charge-Kerr-1}. 
However our result is valid to \emph{all orders} in $\alpha$, not
just the decoupling limit, and at no point have we
imposed the metric equations of motion.
Our proof also generalises to stationary stars as long as the matter
fields do not couple to $\vartheta$ (the matter Lagrangian is
independent of $\vartheta$).  In this case there is no interior
boundary and we get $q = 0$ (see~\cite{YSY} for an earlier approach
based on the generalised-Gauss-Bonnet-Chern theorem; the main advantage of our proof is that it is based on local and covariant quantities).

Our proof can also be adapted to any theory (in any number of spacetime
dimensions) where the dilaton field is linearly coupled to a
topological density \(\mc T\) which depends on the curvature with suitable modifications of \cref{eq:J-Q-defn} with terms of the form \(\delta \mc T/\delta R_{\mu\nu\lambda\rho}\). We can also consider theories where a scalar field is linearly-coupled to a topological density of any \(G\)-bundle where \(G\) is some, possibly non-abelian, group. In this case, the charge contribution takes a form similar to \cref{eq:charge-Wald} with the Riemann tensor replaced by the curvature in the \(G\)-bundle (see \cite{KP-first-law}). As an example we consider briefly a massless
axion field $\varphi$ coupled with strength $g$ to electromagnetism
through the second \emph{Chern character} \cite{BT-book}, via the Lagrangian
\begin{align}
  \df{L}_{\varphi} ={}& \frac{1}{2} (* d\varphi) \wedge d\varphi +
                      \frac{g}{2} \varphi \  \df F \wedge \df F \\
  ={}& \frac{1}{2} (* d\varphi) \wedge d\varphi +
     \frac{g}{2} \varphi \ d(\df F \wedge \df A)
     \,, \nn
\end{align}
where \(\df A\) is the electromagnetic vector potential, i.e.~a
\(U(1)\)-connection, and \(\df F \defn d \df A \) is the field
strength.
For the analogue of \cref{eq:J-Q-defn}, we then have
\begin{subequations}\begin{align}
  \df{\mc J}^{(\varphi)}_X & = \Lie_X (*d\varphi) +
  g \df F \wedge \Lie_X \df A \\
  \df{\mc Q}^{(\varphi)}_X & = X \cdot (*d\varphi) +
  g \df F (X \cdot \df A)
  \,.
\end{align}\end{subequations}
Following through the proof of \cref{thm:main-thm}
under the assumption that \(\df A = \mc O(1/r)\) at spatial infinity,
the axion scalar charge for a black hole is given by
\begin{align}
  q^{(\varphi)} =  g V_{\ms H} Q_m
  \qquad\text{where}\qquad
  Q_m \defn \tfrac{1}{4\pi} \int_B \df F
  \,,
\end{align}
and \(V_{\ms H} \defn K \cdot \df A \vert_B\) is the horizon
potential, which is constant on \(\ms H\) (see \cite{Carter} and
Theorem~1 \cite{KP-first-law}).  Here $Q_{m}$ is the black hole's
magnetic charge,
proportional to the first Chern number of the $U(1)$-bundle over
$B$ \cite{BT-book}.\\

A number of recent investigations~\cite{Doneva:2017bvd, Silva:2017uqg,
  Antoniou:2017acq} focused on a non-linear coupling between the
dilaton and Euler density, replacing $\vartheta\df{\mc E} \to
f(\vartheta) \df{\mc E}$ in the Lagrangian \cref{eq:L-dilaton}
(with a nonlinear function $f(\vartheta)$, the theory no
longer has the shift symmetry $\vartheta \to \vartheta +
\mathrm{const.}$).
These authors pointed out that when $f'(\vartheta)$ vanishes at some
value $\vartheta_{0}$, such a theory admits standard (``no hair'') GR
solutions with a constant dilaton field $\vartheta = \vartheta_{0}$.
However, if $f''(\vartheta_{0})>0$, these solutions can be unstable,
and revert to a stable branch of black hole solutions with
dilaton hair.

Analyzing this coupling, we again have \cref{eq:J-dQ} with
\be\begin{split}
   \df E_\vartheta & =  d * d\vartheta + \tfrac{\alpha}{8} f'(\vartheta) \df{\mc E} \\
    \df{\mc Q}_X & = X \cdot d\vartheta + \tfrac{\alpha}{4} f'(\vartheta) \epsilon_{abcd} \df R^{ab} (X \cdot \df\omega^{cd}) \\
    \df{\mc J}_X & = \Lie_X (*d\vartheta) + \tfrac{\alpha}{4} f'(\vartheta) \epsilon_{abcd} \df R^{ab} \wedge \Lie_X \df\omega^{cd} \\
    &\quad + \tfrac{\alpha}{4} f''(\vartheta) \epsilon_{abcd} \df R^{ab} \wedge d \vartheta~ (X \cdot \df\omega^{cd})
    \,.
\end{split}\ee 
Note that in the nonlinear case, \(\df{\mc J}_X\) is still conserved, but
\(\df{\mc J}_X \neq 0\) even for a symmetry \(X^\mu\) of the solution,
because there is no more symmetry under the shift $\vartheta \to
\vartheta + \mathrm{const}$.
Therefore, a bulk integral term will remain in the computation of the charge.
A repetition of our proof gives the dilaton charge
\be\begin{split}
  \label{eq:charge-nonlinear-f}
    q & = \tfrac{\alpha}{8\pi} \kappa \int_B \df\varepsilon_2 f'(\vartheta) R_2 \\
    &\quad - \tfrac{\alpha}{8\pi}\int_\Sigma \df\varepsilon_3 f''(\vartheta) u_\mu ({}^*R^*)^{\mu\nu\lambda\rho} \nabla_\nu \vartheta \nabla_\lambda K_\rho
    \,,
\end{split}\ee
where \(\df\varepsilon_3\) is the induced volume element and \(u^\mu\)
is the unit timelike normal to the Cauchy surface \(\Sigma\).

In \cref{eq:charge-nonlinear-f}, we can easily see the difference
between the linearly coupled case and the non-linear case.  Because of
the lack of shift symmetry in the nonlinear case, the bulk term
remains (i.e.~$f''(\vartheta)\neq 0$),
and thus the dilaton charge depends on the metric and dilaton
solutions throughout the entire spacetime.
We no longer get a relation between quantities evaluated
purely on the boundaries.
However, if the dilaton field has a small variation throughout
spacetime, it may be possible to expand the theory~\cite{YSY} around
some typical value $\vartheta_{1}\neq \vartheta_{0}$, where
$f'(\vartheta_{1})\neq 0$.  Then if we expand the coupling function to
linear order around \(\vartheta_1\), we recover the shift-symmetric,
linearly-coupled theory.

\acknowledgments
We would like to thank
B\'eatrice~Bonga and Robert M.~Wald
for useful conversations.
We also thank David Garfinkle for comments on an earlier draft of the paper. 
K.P. is supported in part by the NSF grants PHY--1404105 and PHY--1707800 to Cornell University.
L.C.S. acknowledges the support of NSF grant PHY--1404569 and the support
of the Brinson Foundation.
Some calculations used the computer algebra
system \textsc{Mathematica}~\cite{Mathematica}, in combination with the
\textsc{xAct/xTensor}
suite~\cite{JMM:xAct,MARTINGARCIA2008597}.

\appendix

\section{Short primer on tetrads and spin connection}
\label{sec:tetrads}

In this appendix we give a short introduction to tetrads and connections. We will use the language of principal bundles, for the details of which we refer the reader to the classic treatment\footnote{Note that these references may use different conventions when converting differential forms to an index notation.} of \cite{KN-book1, CDD-book, CD-book} (see also the appendix of \cite{KP-first-law}).

On spacetime, the \emph{oriented orthonormal tetrads} \((\df e^a)_\mu \equiv e^a_\mu\) are defined by
\be
    g_{\mu\nu} = \eta_{ab} e^a_\mu e^b_\nu \eqsp \varepsilon_{\mu\nu\lambda\rho} = \epsilon_{abcd} e^a_\mu e^b_\nu e^c_\lambda e^d_\rho
    \,,
\ee
where \(\eta_{ab} = {\rm diag}(-1,1,1,1)\) and \(\epsilon_{abcd}\) are the metric and orientation in \(\bb R^4\) with \(\epsilon_{0123} = 1\). The ``inverse'' tetrads \(e_a^\mu\) satisfy
\be
    e_a^\mu e^a_\nu = \delta^\mu_\nu \eqsp e_b^\mu e^a_\mu = \delta^a_b
    \,.
\ee
The torsion-free \emph{spin connection} \(\df\omega^a{}_b\) is given by
\be
    (\df \omega^a{}_b)_\mu = e^a_\nu \nabla_\mu e_b^\nu   
    \,.
\ee

Given a metric \(g_{\mu\nu}\), the tetrads and the spin connection are only determined up to a local Lorentz transformation \(\Lambda^a{}_b(x)\) which depends on the the point \(x\) in spacetime
\be
    \df e^a \mapsto \Lambda^a{}_b \df e^b \eqsp \df\omega^a{}_b \mapsto \Lambda^a{}_c \df\omega^c{}_d (\Lambda^{-1})^d{}_b + \Lambda^a{}_c d (\Lambda^{-1})^c{}_b
    \,.
\ee
We note that the connection transforms non-covariantly. Due to this ``internal gauge freedom'' it is more natural to treat these as fields on a principal bundle with structure group given by the Lorentz group.

Differential forms on spacetime which are covariant under Lorentz transformations are represented by \emph{horizontal} differential forms on the bundle. The connection is represented by a Lie algebra-valued \(1\)-form (which by definition is not horizontal). The connection is then determined uniquely by the torsion-free condition
\be
    0 = D\df e^a = d \df e^a + \df \omega^a{}_b \wedge \df e^b
    \,,
\ee
where \(D\) is the \emph{covariant exterior derivative} defined by \(\df \omega^a{}_b\). The curvature \(2\)-form is defined by the horizontal form
\be\label{eq:R-defn}
    \df R^a{}_b \defn D\df \omega^a{}_b = d \df\omega^a{}_b + \df \omega^a{}_c \wedge \df\omega^c{}_b 
    \,,
\ee
Since the curvature is horizontal it represents a covariant form on spacetime related to the Riemann tensor as
\be
    (\df R^a{}_b)_{\mu\nu} \equiv R^\lambda{}_{\rho\mu\nu} e^a_\lambda e_b^\rho
    \,,
\ee
and the Bianchi identity reads
\be\label{eq:bianchi}
    D\df R^{ab} = 0 \implies \nabla_{[\sigma} R^{\lambda\rho}{}_{\mu\nu]} = 0 \,.
\ee

The Lie derivative of the tetrads and connection with respect to vector fields on the bundle is
\begin{subequations}\begin{align}
    \Lie_X \df e^a & = D(X \cdot \df e^a) - (X \cdot \df\omega^a{}_b) \df e^b \label{eq:Lie-e}\\
    \Lie_X \df\omega^a{}_b & = X \cdot \df R^a{}_b + D (X \cdot \df\omega^a{}_b) \label{eq:Lie-conn}
\end{align}\end{subequations} 
From the spacetime point of view, this encodes the fact that Lie derivatives of the tetrads and connection are only defined up to a local Lorentz transformation (encoded in the vertical part of the bundle vector field).

It can be shown (see Lemma A.2 \cite{KP-first-law}) that a bundle vector field which preserves the tetrads, \(\Lie_X \df e^a = 0\), projects to a Killing field \(X^\mu\) of the metric on spacetime, and further satisfies
\be\label{eq:symm-cond}
    X \cdot \df\omega^{ab} = - e^a_\mu e^b_\nu \cd^\mu X^\nu 
    \,,
\ee
where the left-hand-side is computed as a function on the bundle.

\bibliographystyle{JHEP}
\bibliography{dilaton-charge}
\end{document}